\documentstyle[aps,preprint]{revtex}
\begin{document}
\preprint{UVA-INPP-97/1}

\title{Off-shell Corrections and Moments of the Deep Inelastic 
Nuclear Structure Functions.}

\author{Christopher D. Cothran $^{*}$, Donal B. Day $^{*}$  
and Simonetta Liuti$^{* \dagger}$}

\address{$^{*}$
Institute of Nuclear and Particle Physics, University of
Virginia.\\Charlottesville, Virginia 22901, USA. \\
$^{\dagger}$
INFN, Sezione Sanit\`a. Physics Laboratory, Istituto Superiore
di Sanit\`a.\\ Viale Regina Elena, 299. I-00161 Rome, Italy.}
\maketitle

\begin{abstract}
We present an improved method of handling off-shell effects in deep 
inelastic nuclear scattering.  With a firm understanding of the effects of the 
nuclear wavefunction, including these off-shell corrections as well as binding 
effects and nucleon-nucleon correlations, we can begin to examine the role of 
QCD in nuclei through an analysis of the moments of the nuclear 
inelastic structure function $F_2^A$.  Our analysis is 
aimed on one side at extracting the $Q^2$ dependence of the 
moments of the nucleon structure function by using the recent 
high $x$ world Iron data and by properly subtracting nuclear effects
from the perturbative contribution.  
On the other side we compare quantitatively the behavior of the extracted 
moments with a simple $O(1/Q^2)$ phenomenological form and we determine the
mass term for this parametrization. 
\end{abstract}
\narrowtext
The discovery of nuclear effects in Deep Inelastic Scattering (DIS) 
ultimately represents the fact that the 
quark and gluon structure of a {\it free} nucleon is different from that  
of a {\it bound} one.
Nuclear effects have been found systematically in a number of 
high 
energy processes using both leptonic and hadronic probes
(see {\it e.g.} \cite{Arne}
and \cite{GeeTho} for recent reviews and references).
Results are most often presented in the form of a ratio, $R_A$, of  
the nuclear cross section per nucleon
to that of the free nucleon, where $R_A =1$ within error bars means 
absence of nuclear effects.     
%
%
%
%

%
%
In this Letter we concentrate on the $Q^2$ dependence of DIS in the 
intermediate and high Bjorken $x$ regions ($x>0.3$). 
Our aim is twofold: ${\bf 1.}$ On one side we investigate in detail the 
struck particle's off-shell effects which, within a class of models 
based on conventional degrees of freedom, can be considered as
the major source of modifications of the $Q^2$ dependence
of the bound nucleon's structure function. 
In this context we make the observation that the proper framework to 
investigate off-shell 
effects in DIS from nuclei has to be sought within a relativistic theory 
and with a definite prescription for going beyond the     
Impulse Approximation (IA). As a side result, we show how one needs 
to be very careful about using ad hoc prescriptions.
${\bf 2.}$ On the other side we construct the quantities needed in a QCD 
analysis of the data, namely the (normalized)  
moments of the nuclear structure functions.
We show that the off-shell effects investigated in the first part of the paper 
have a very small impact on the moments or that, in other 
words, nuclear effects can be factored out almost 
exactly. This allows us to extract the coefficients of 
the Higher Twist (HT) terms of the nucleon structure functions 
more accurately than previous determinations (\cite{PenRos,AbbBar} 
and references therein).  
 
%
%

${\bf 1.}$ 
While for the free nucleon it has been possible to address this question 
entirely within QCD \cite{EPP,DGP,EFP},
a natural complication arises for the nucleus \cite{Coh}.
On a phenomenological level we see that there are now  
two types of
constituents, both the struck quark 
and the parent nuclear constituent, {\it e.g.} a nucleon, 
that can be off-shell.
%
%
As we will show, the quark's tranverse momentum and its virtuality are 
related differently in a free and in a bound nucleon, respectively and 
this affects their relative structure functions which, within a parton model 
depend on the transverse degrees of freedom (or on the virtuality) through
the particles propagators.      
A careful treatment of off-shell effects along these lines, turns out to 
be crucial and in fact {\em ad hoc} procedures, 
such as the popular 
DeForest prescription \cite{DeF}, fail consistency in the DIS region, 
{\it e.g.} by predicting a 
so far unobserved pattern of $x$-scaling violations.

One can describe the DIS process for nuclear targets with a 
``double'' IA,
according to which the virtual photon scatters off a 
quark that in turn belongs to a nuclear constituent ({\em e.g.} a nucleon).  
IA is expected to work 
at the $Q^2$ values of interest ($Q^2 \geq 2 \, {\rm GeV}^2 $) and at 
$x$ not in the proximity of 0 and 1.    
By using a description  
in which the struck quark and nucleon transverse degrees of 
freedom  are decoupled
(see {\it e.g.} Ref.\cite{Bir}), 
the nuclear structure function can be written as a convolution formula 
in the {\em longitudinal} variables
\begin{equation}
F_2^A(x)= \int_x^A dz f_A(z) F_2^N \left( \frac{x}{z} \right),
\label{conv}
\end{equation}
where $z=(k_0-k_3)/M \nu$ 
is the light-cone (LC) momentum fraction carried by the nucleon, 
$k_o$ being the bound nucleon's energy which is obtained from energy 
conservation at the nuclear vertex: $k_0=M_A - \sqrt{M_{A-1}^{* \, 2}+k^2}$
and $k_3 \equiv {\bf k} \cdot {\bf q}$.
\footnote
{Note $M^*_{A-1}= M_{A-1} + E_{A-1}^*$ where $E_{A-1}^*$ is the 
excitation energy of the outgoing $A-1$ nucleons system which is 
related to the nucleon's removal energy, $E$,  by $E=E_{min} + E_{A-1}^*$, 
$E_{min}=M_N + M_{A-1}-M_A$.}
$f_A(z)$ is the nucleon's LC momentum fraction distribution, 
\footnote 
{$f_A(z)$ properly includes the {\em flux factor} in its definition 
and it is normalized so as to satisfy baryon number 
conservation \cite{Bir,Liu1}.}  
and $F_2^N$ is the nucleon structure function.
Calculations based on Eq.(\ref{conv}) reproduce   
the slope of $R_A$ at intermediate $x$ ($0.3 \leq x \leq 0.55$), 
provided  
a description of nuclear dynamics which properly 
accounts for Nucleon-Nucleon (NN) correlations   
in the nucleon momentum and removal energy spectra is adopted \cite{Liu1} 
(see also 
\cite{GeeTho,Bir}). 
%
%

%
%
Theoretical predictions fall short
of the data outside the interval $ 0.3 \leq x \leq 0.55$. 
%
%
%
%
It should be noted at this point that we have focused on the {\em nucleon}
components only and that indeed Eq.(\ref{conv}) does not satisfy momentum
conservation. This problem is generally solved by introducing scattering  
from additional nuclear pions  \cite{GeeTho,Roberts} which while restoring
momentum conservation, also modifies the slope of the data at low $x$ (at
very low $x$ nuclear shadowing eventually becomes dominant). 
Additional pions are not going to be of immediate relevance to our thesis
which is to study the $Q^2$ dependence of the moments of the non-singlet 
structure functions, for which the high $x$ behavior is important.    
   

The next step is therefore 
to investigate whether the poor agreement with the data at high $x$ 
is 
an indicator
that IA is a 
poor approximation at 
$Q^2 > 2 \, {\rm GeV}^2$. 
%
%
In other words the effect of parton interactions can be sizeable 
and it can manifest itself differently inside a nucleus.   
We begin investigating this problem by defining the role 
of the transverse degrees of freedom and of 
the struck parton's virtuality (or off-shellness)
in a nucleus. 
%
%

%
%

Notice that now both 
the struck parton and the parent nucleon are 
off-shell and the struck parton and spectator interactions can occur in 
more ways than in a free proton. 
With the intent of investigating this problem systematically and in 
a parton language suitable to nuclear targets, we start by comparing the 
on-shell and off-shell nucleon structure functions.   
 
The structure functions are extracted from the cross
section $\sigma(\gamma^* P \rightarrow k' X)$
($k'$ and $X$ being the final quark and the spectator system). 
By using an infinite 
momentum frame parametrization of the particles four-momenta and 
Old-Fashioned-Perturbation-Theory  (OFPT)
one obtains   
\begin{equation}
F_2^{ON}(x) = \frac{x^2}{1-x} \int \frac{d^2 {\bf p_T}}{2(2 \pi)^3} 
\left[ \frac{\phi(S)}{x^2(S-M_N^2)} \right]^2,  
\label{f2free}
\end{equation}
where the struck parton's momentum components 
are $p_\mu \equiv(xP + m^2/2xP; {\bf p_T}, xP)$, the spectators  
ones are $P_X \equiv 
((1-x)P + m_X^2/2(1-x)P; -{\bf p_T},(1-x)P)$, a sum over the $m_X$ mass 
spectrum is implied in Eq.(\ref{f2free}),
\footnote{The $m_X$ distribution really determines the behavior
of $F_2$ only at $x \approx 0$ anyway.}    
$S\equiv S(x,p_T^2)=p_T^2/x(1-x) + m_X^2/(1-x) + m^2/x$,  
$\phi(S)$ is the vertex function, 
and finally      
we are assuming scalar particles because our argument  
will be essentially kinematic (Eq.(\ref{f2free}) can be easily 
generalized to spin 1/2 constituents). 
Notice also that the final result is Lorentz invariant and that the 
integration in Eq.(\ref{f2free}) can be directly transformed into 
an integral over $S$, thus transfering all the $x$ dependence into 
the lower limit of integration: $S_{min}= m_X^2/(1-x) + m^2/x$.

By considering the same graph inside a nucleus, one obtains for 
the off-shell nucleon structure function  
\begin{equation}
F_2^{OFF} (y,k^2_T) = \frac{y^2}{1-y} \int   
\frac{d^2 {\bf p_T}}{2 (2\pi)^3} \left[ \frac{\phi_A
(\widetilde{S})}{y^2(\widetilde{S}-M_N^2)}
\right]^2. 
\label{f2off}
\end{equation}
where now $\widetilde{S}\equiv S(y,a_T^2)=a_T^2/[y(1-y)] + m_X^2/(1-x) 
+ m^2/y$, 
with
${\bf a_T}= {\bf p_T} - y {\bf k_T}$, 
$y=x/z$, because of Fermi motion (the struck parton and the spectators have 
now
perpendicular momenta: ${\bf p_T}$ and ${\bf k_T} - {\bf p_T}$, respectively,
$k_T$ being the nucleon transverse momentum). 
Two important observations emerge from the comparison of
Eqs.(\ref{f2free}) and (\ref{f2off}).  
First of all note the role 
played by the nucleon's $k_T$: while $\langle z \rangle \approx 1$ and 
therefore rescaling effects in the longitudinal variables are predicted to 
be small, $k_T$ and $p_T$ are similar in magnitude and we expect this to
affect the derivation of the next to leading twist correlation functions 
in a nucleus \cite{CotLiu2}. 
Secondly one obtains $\langle k_T^2 \rangle/M_N^2$ modifications to the
naive convolution formula (\ref{conv}), that can
be calculated by relating  
$F_2^{ON}$ and $F_2^{OFF}$ 
by a shift of their longitudinal variables:
$F_2^{ON}(y^\prime) = J(y,y') F_2^{OFF}(y\equiv x/z,k_T)$, where 
$y' \sim y \times (k_T^2 + m_X^2)/(y k_T^2 + m_X^2)$, where
$J(y',y)$ is a kinematical factor. 
This relation between $F_2^{ON}$ and $F_2^{OFF}$ 
is obtained by transforming the integral 
over ${\bf k_T}$ into an integral over $\widetilde{S}$ 
and by observing that the lower limit of integration, 
$\widetilde{S}_{min} =  m_X^2/(1-y) + m^2/y + y p_T^2/(1-y)$, 
is related to the free nucleon $S_{min}$, calculated above 
by a shift of momentum and mass (this technique has been 
applied often in OFPT, see Ref.\cite{Close} and references therein). It is 
valid within the hypothesis that
the vertex function is not modified inside the nuclear medium, 
($\phi_A \equiv \phi$), which is reasonable given the ``locality'' of the
electron-nucleus interaction. 
The convolution integral that properly takes into account the transverse
degrees of freedom (and the parton virtuality) is given by
\begin{equation}
F^A_2(x) = 2 \pi M_N
\int_x^\infty dz   \, z
\int ^\infty _{k_{min}(\langle E \rangle _A,z)} dk\,k\,
n_A(k) F^{ON}_2[y'(k_T^2)], 
\label{off}
\end{equation}
where $k  \equiv \sqrt{[M_N(1-z)-E]^2 + {\bf k_T}^2}$, $n_A(k)$
is the 
nucleon momentum distribution \cite{Liu1}, 
and both the Jacobian,
$2 \pi  M_N$, and the lower limit 
of integration, $k_{min} = \mid M_N(1-z)- \langle E \rangle _A \mid$,  
come from the trasformation of variables
$(k_\parallel, {\bf k_T}) \rightarrow (z,k)$.
Eq.(\ref{off}) satisfies, like Eq.(\ref{conv}), baryon number conservation 
but not momentum conservation because, as mentioned before, it only involves
nucleons while it is well known that momentum can also 
be carried by nuclear pions. Eq.(\ref{off}) predicts a large EMC effect than
Eq.(\ref{conv}) essentially because  $F^{ON}_2(y') < F^{ON}_2(y)$, being
$ y' > y$.      

Therefore we conclude that
in DIS scattering from a nucleus, the nucleon Fermi motion
modifies the relationship between 
the struck parton virtuality and its transverse momentum, with respect to 
that of a free proton. 
It is well known that at 
high $x$ the electron-nucleus cross section is sensitive
to the high momentum components, {\it i.e.} to coherent effects arising 
from NN correlations that, as seen in Eqs.(\ref{f2off}) and (\ref{off}), 
can affect both the parton $p_T$ and $x$ distributions.
The physics at high $x$ deals with the interplay between these 
coherent effects and the pure QCD  
interactions between the struck and spectator partons in the nucleus.
This perhaps subtle point 
is the main result of the present paper which is intended to prepare the 
ground for our further investigation of higher twists. 
Our approach is substantially different from some commonly used ones 
that are based on a longitudinal convolution formula (Eq.(\ref{conv})) 
while taking care of the off shell nucleon through {\em ad hoc} 
prescriptions. Such approaches are not only 
conceptually wrong, based on our previous discussion, 
but we show that they predict a pattern of $x$-scaling violations
that is not observed in the data. In particular, we discuss   
the most popular model, obtained by extending to DIS   
the  ``De Forest's prescription'' \cite{DeF} which  
was originally constructed for electron-nucleus 
Quasi-Elastic (QE) scattering and 
first applied to nuclear DIS in Ref.\cite{Sau} (see also \cite{Ingo} for
nuclear matter). 
The underlying philosophy is that lacking 
any knowledge on the form of the operators connecting off-shell 
nucleon states, one should 
reevaluate the kinematics 
according to:
\begin{mathletters}
\begin{eqnarray}
k_o \equiv M_N - E - T_{A-1} & \rightarrow & 
\bar{k}_o = \sqrt{ {\bf k}^2 + M_N^2}, \\
\label{pa}
\nu & \rightarrow & \bar{\nu} = \nu - \left(\bar{k}_o - k_o \right),  
\label{pb}
\end{eqnarray} 
\end{mathletters}
where $T_{A-1}= k^2/2M_{A-1}$ 
is the $A-1$ system's recoil energy. 
$k_o$ is the off-shell nucleon energy whereas $\bar{k}_o$ is DeForest's 
proposal for its on-shell substitute.  
This method essentially reduces to putting the initial nucleon back on shell,
at the expense of modifying the energy dependence of the four-momentum 
transfer, $q_\mu$.  Equivalently, the ambiguity due to the interaction 
involving an off-shell nucleon current is solved by ``rescaling'' the $Q^2$ 
dependence of the 
nucleon structure functions.
%
By implementing 
Eqs.(\ref{pa}, \ref{pb}) in the calculation of the inelastic cross section, 
one also obtains a breakdown of the convolution formula because 
of the modifications in the $Q^2$ which is shared by the struck 
nucleon \cite{Sau}.
%
%

${\bf 2.}$ 
Since our final objective is to understand QCD in nuclei, 
in the last part of this paper we concentrate on the moments of 
the nuclear  structure functions. 
It is a well known consequence of the IA with no off-shell effects that 
the nuclear moments factorize 
into the ($Q^2$-dependent) nucleon moments and the ($Q^2$-independent) 
moments of the nuclear light cone distribution. In what follows 
we show our theoretical expectations based on the previous
discussion of off-shell effects along with 
our extraction
of nuclear moments from current world 
experimental data.
From our analysis we are able to determine the 
coefficient of the $O(1/Q^2)$ correction and to compare 
it with the results of the quantitative analysis of Ref.\cite{PenRos} 
and \cite{AbbBar} which were
obtained from data extending up to a lower value of $x$. 

The Cornwall-Norton moments of the nuclear structure functions read
\begin{equation}
M_n^A(Q^2) = \int _0^A dx x^{n-2} F_2^A(x,Q^2) . 
\end{equation}
If $F_2^A$ is written in terms of a longitudinal convolution formula
the following factorization holds \cite{Roberts}
\begin{eqnarray}
M_n^A(Q^2) & = & \left[ \int_0^A dz f_A(z) z^{n-1} \right] 
\,  \int _0^A dx x^{n-2} F_2^N(x,Q^2)   \equiv 
\nonumber \\  
& \equiv &  {\cal M} _n^A \,  M_n^N(Q^2), 
\label{mom1}
\end{eqnarray}
which is such that the ratio $M_n^A(Q^2)/M_n^N(Q^2)$ is independent 
of $Q^2$. 
A breakdown of the longitudinal convolution formula translates 
into a breakdown of such a factorization.
We examined two different mechanisms deriving from 
off-shell effects, namely: {\it (i)} the presence of transverse 
degrees of freedom in the present 
approach, and {\it (ii)} 
the modification of the way $Q^2$ is shared by the struck nucleon
according to \cite{DeF,Sau}.
%
%
Our results are presented in Fig.1 for $n=3-5$. 
All theoretical results were obtained by
using the same realistic spectral function which includes NN correlations
\cite{CDL}.
The experimental data 
are for $^{56}Fe$; they have been obtained from an analysis of world 
data to be discussed later. One can see that while the convolution formula
predicts no $Q^2$ dependence at all, prescription {\it (i)} shows a very 
slight breaking of factorization which is evident from the small slope 
in $Q^2$ and prescription {\it (ii)} seems to display an unphysical wiggle
and a steeper slope, due to the way the $Q^2$ is rescaled according to 
Eqs.({\ref{pa},\ref{pb}). The existing 
data are not accurate enough to detect any slope in the given $Q^2$ range.
%
%

In order to optimize  
the extraction of the $Q^2$ dependence of 
$M_n^A$ we then investigated the following quantity: 
${\cal R}_n^A = 
\left[ M_n^A(Q^2)/M_n^A(Q_o^2) \right] ^ {-1/d_n}$. In the free nucleon
case and 
for the non singlet (NS) part of the structure function, the  
LO perturbative QCD, prediction for this quantity is
%
%
\begin{equation}
\left[ \frac{M_n^{N}(Q^2)}{M_n^{N}(Q_o^2)} \right] ^ {-1/d_n}  
=  \frac{\alpha_S(Q_o^2)}{\alpha_S(Q^2)} \equiv 
C \; \left( ln \, Q^2 - ln \, \Lambda_{QCD}^2 \right)   
\label{Penni}
\end{equation}
where $d_n = \gamma_n / 2 \beta_o$, $\gamma_n$ are the NS
anomalous dimensions, and $\beta_o = 11 - {2}{3} N_f$.
$Q^2_o$ is an arbitrary scale to be chosen conveniently. 
The motivation for normalizing the nuclear moments at a scale $Q_o^2$
can be understood by looking at the behavior of the curves in Fig.1,
where the different treatments of off-shell effects are shown to 
produce overall depletions of different size. In the ratio ${\cal R}_n^A$
such differences are leveled and one can concentrate on the 
$Q^2$-dependence only.   
Having in mind Eq.(\ref{mom1}), in Fig.2 we plotted ${\cal R}_n^A$ 
vs. $ln Q^2$. 
For a free nucleon any deviation from
a straight line can be considered as a signal of higher twist contributions,
since it has been shown \cite{PenRos} that NLO terms are negligible.
For a bound nucleon one has the initial problem of distingushing 
the off-shell effects originating from the nucleon transverse degrees of 
freedom, from ``pure QCD'' higher twist 
effects.  
The data points shown are extracted from the results of experiments 
at CERN (NA2) \cite{NA2} and SLAC (E139,NE18) \cite{E139,NE18}.
To combine these different sets, most of the data needed to be 
interpolated to common values of $Q^2$ to generate measured 
$F_2^A$ curves.  The NE18 \cite{NE18} 
data provided the high $x$ values of $F_2^A$; 
however, the QE contribution was large and its removal by a 
calculation \cite{CDL} 
introduced additional systematic error. The extrapolation 
to values of $x$ beyond the region of the data also introduced error.  
At low $x$, the error is small and becomes 
negligible for moments $n>2$.  At high $x$ the extrapolation error is 
significant except where NE18 data is available. This error was estimated by 
looking at the range of values produced by different extrapolation methods.  
For all moments the extrapolation error is at least 
as large as the statistical uncertainty, and becomes dominant for large $n$.

We compare data with our results for the convolution formula 
and our off-shell prescription. The normalization point is at 
$Q_o^2= 12.5 \, {\rm GeV}^2$. We present 
results for $n \geq 3$ because here the contribution 
of the singlet part of the structure function becomes negligible. 
%
The off-shell curve deviates 
very little from the free nucleon one Eq.(\ref{Penni}) 
(they are indistinguishable in the figure). Therefore   
we conclude that the off-shell effects
considered here, {\it i.e.} the ones that derive from the modifications
to the struck quark $p_T$ in a bound nucleon, 
have been properly subtracted     
through the use of the normalized ratios shown in Fig.2.   

We can now look for the presence of higher twist terms in the data.  
In order to address this problem quantitatively, 
we started from considering the simple parametrization \cite{DGP,PenRos}  
\begin{equation}
M_n^A(Q^2)= M_n^{pQCD}(Q^2) \left( 1 + \frac{n \tau^2}{Q^2} \right).
\label{HT}   
\end{equation}
where $M_n^{pQCD}(Q^2)$ is given by Eq.(\ref{Penni}) and $\tau^2$ is the mass
parameter to be determined. We assume initially that the $O(1/Q^2)$ 
correction is not A-dependent. Our results are presented in Fig.3 for 
a fixed range of values of $\Lambda_{QCD} = 0.200 - 0.260 \, {\rm MeV}$.
The values displayed show the sensitivity to the low $Q^2$ region: 
results change dramatically by excluding the point at $Q^2= 2 \, {\rm GeV}^2$
(diamonds), and by including it (stars) where after $n=5$ ({\it i.e.} where
high $x$ also dominates), the 
validity of Eq.(\ref{HT}) becomes doubtful.          
Our result is somewhat smaller than in the previous extraction,
$\tau^2=(0.8 \pm 0.2 \, {\rm GeV})^2$, \cite{AbbBar}, 
but it is consistent with 
the original estimate,
$\tau^2=(0.4 \, {\rm GeV})^2$, \cite{DGP};  
and with Ref.\cite{PenRos} 
(see area between the dotted lines in Fig.3).    
%

In conclusion, we have shown that the complications 
of kinematical origin due to nuclear effects and nucleon off-shellness   
contribute a small, if not negligible, modification
to the $Q^2$ dependence of the moments of the nuclear structure function 
$F_2^A$. Therefore the consequences of QCD in nuclei can be examined in 
detail. In particular, we successfully extracted the coefficient of the 
$O(1/Q^2)$ power corrections.  As these effects are 
prominent in the region of low $Q^2$ and large $x$, 
we anticipate the forthcoming Jefferson 
Laboratory experiments \cite{Donal,Thia} to
significantly reduce the experimental uncertainties in the extraction of the 
moments of $F_2^A$.


We thank Ingo Sick for raising interesting questions. 

This work was partly
funded by  the U.S. Department of Energy (D.O.E.) $\#DE-FG0586ER40261$.


\begin{figure}
\caption{Ratio of the moments for $^{56}Fe$ to the nucleon ones. The experimental data
extracted in this paper are compared to: (a) the convolution formula; 
(b) the off-shell model described in this paper; (c) the De Forest 
prescription.}  
\end{figure}

\begin{figure}
\caption{Pennington-Ross plots for the world data on $F_2^A$. 
Solid curves:
LO theoretical results using the convolution formula; 
The singlet
contribution is shown for $n=3$ (short dashed line) and it is negligible 
at higher $n$. The off-shell effects are shown for n=4 (short dashed line)
as an expample since they are of the same magnitude, and therefore 
negligible, for the other values of $n$. 
Dotted curves: expectation for the ``simple'' higher twist contribution 
(Eq.(\protect\ref{HT})), 
calculated with $\tau^2$ from our experimental fit.}  
\end{figure}

\begin{figure}
\caption{The mass parameter (squared), $\tau^2$, which regulates the simple 
twist-4 expectation of Eq.(\protect\ref{HT}). The diamonds are obtained 
from our fits to the ${56}Fe$ data, excluding the point at 
$Q^2= 2 {\rm GeV}^2$; the stars include $Q^2= 2 {\rm GeV}^2$;
the two dotted lines show the range of the predictions by 
Ref.\protect\cite{PenRos}. The value of $\tau^2$ originally obtained 
by de Rujula et al., was $0.16 {\rm GeV}^2$, which is still in 
qualitative agreement with our results.}      
\end{figure}


\begin{references}
\bibitem{Arne} M.Arneodo, Phys. Rep.{\bf 240} (1994) 301.
\bibitem{GeeTho} D.F.Geesaman, K.Saito and A.W.Thomas, 
Ann. Rev. Nucl. Part. Sci. {\bf 45} (1995) 337.
\bibitem{EPP} R.K.Ellis, G.Parisi, R.Petronzio, 
Phys. Lett. {\bf 64B} (1976) 97.
\bibitem{DGP} A. De Rujula, H. Georgi and H.D. Politzer, 
Phys. Rev. {\bf D15} (1977) 2495; Ann. of Phys. {\bf 103} (1977) 315.
\bibitem{EFP} R.K.Ellis, W.Furmanski and R.Petronzio, Nucl. Phys. {\bf B212}
(1983) 29. 
\bibitem{Coh} T.D. Cohen, Talk delivered at the {\it Workshop on the Deep
Inelastic Structure of Nuclei held at Jefferson Lab}, December 1996, 
Eds. S. Liuti and W. Melnitchouk, p.77. 
\bibitem{DeF} T. DeForest, Nucl. Phys. {\bf A 392} (1983) 232.
\bibitem{PenRos} M.R. Pennington and G.G. Ross, 
Nucl. Phys. {\bf B179} (1981) 324. 
\bibitem{AbbBar} L.F. Abbott and R.M. Barnett, 
Ann. Phys. {\bf 125} (1980) 276. 
\bibitem{Bir} B.L. Birbrair, E.M. Levin and A.G. Shuvaev, Nucl. Phys. 
{\bf A491} (1989) 618.
\bibitem{Liu1} C. Ciofi degli Atti and S. Liuti,
Phys. Lett. {\bf 225B} (1989) 215;
Phys. Rev. {\bf C41} (1990) 1100.
\bibitem{Close} F.E. Close, {\it ``An Introduction to Quarks and Partons//},
Academic Press, Fifth Printing (1989) pp.199-205. 
\bibitem{CotLiu2} C. Cothran and S. Liuti, in preparation. 
\bibitem{Sau} H. Meier-Hajduk, U. Oelfke and P.U. Sauer, Nucl. Phys
{\bf A500} (1991) 637 and {\it ibid} {\bf A499} (1989) 637.
\bibitem{Ingo} O.Benhar, V. Pandaripande and I.Sick, Phys. Lett. 
{\bf B410} (1997) 79.
\bibitem{Roberts} R.G. Roberts, {\it ``The Structure of the Proton''}, 
Cambridge University Press (1990).
\bibitem{CDL} C. Ciofi degli Atti, D.B. Day and S. Liuti, 
Phys. Rev. {\bf C46} (1992) 1045.
\bibitem{NA2} J.J. Aubert et al., Phys. Lett., {\bf 123B} (1983) 275.
\bibitem{E139} J. Gomez et al., Phys. Rev. {\bf D49} (1994) 4348.  
\bibitem{NE18} J. Arrington et al., Phys. Rev. {\bf C53} (1996) 2248.  
\bibitem{Donal} TJNAF Experiment E89008, D.B. Day and B.W. Filippone 
spokespersons.
\bibitem{Thia} C. Keppel, private communication.
\end{references}
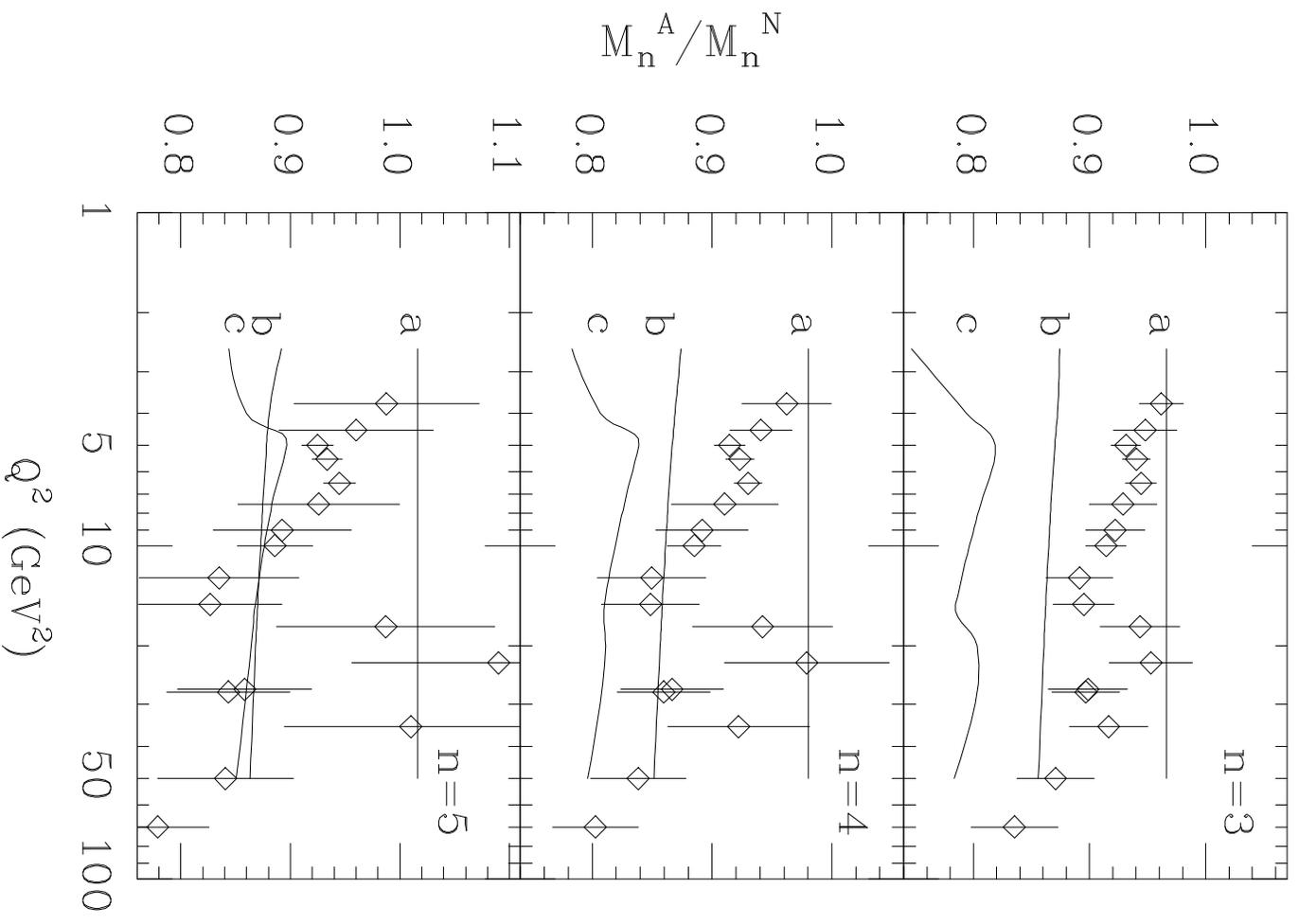
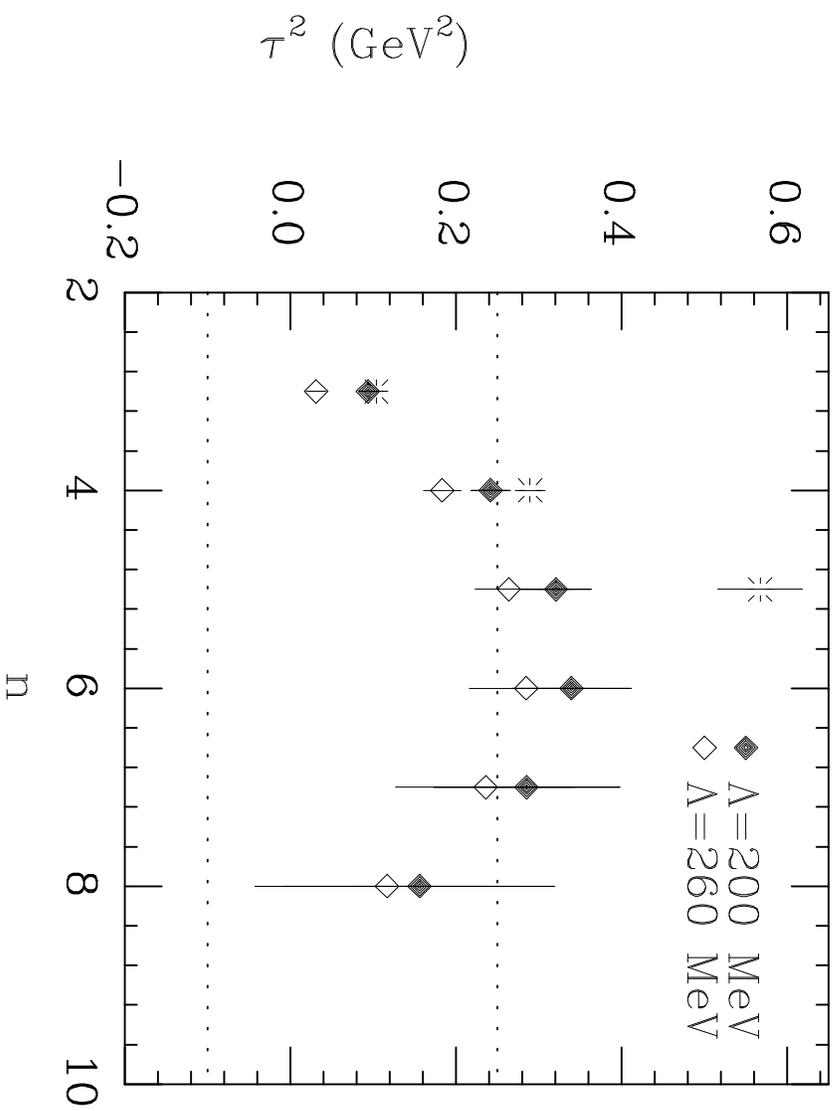
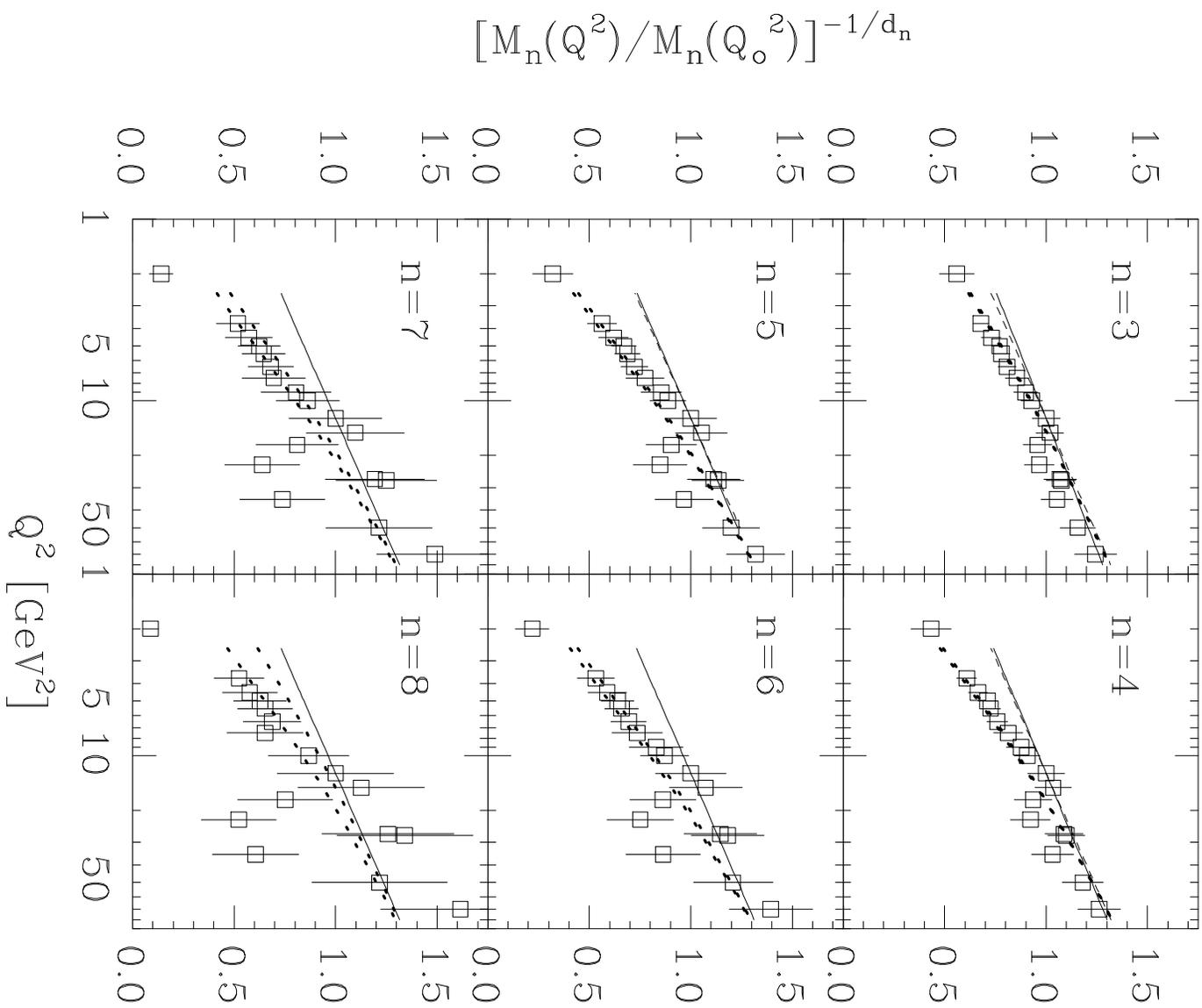
\end{document}